\documentstyle[12pt,fleqn]{article}
\newcommand{\eqnsection}{
\renewcommand{\theequation}{\thesection.\arabic{equation}}
\makeatletter
\csname $addtoreset\endcsname
\makeatother}
\input{dlat}
\begin{document}
\eqnsection
\title{ Derivative and higher order extensions of Davey-Stewartson
equation from matrix KP hierarchy
  }
\author{Anjan Kundu$^\dagger$  and  Walter Strampp
 \\
Fachbereich 17--Mathematik/ Informatik \\
GH--Universit\"at Kassel \\
Holl\"andische Str. 36 \\ 34109 Kassel, Germany
}
\date{}
\maketitle
\begin{abstract}
%------------------------------------------------------------
It is shown that the matrix KP hierarchy can yield new integrable
equations in $(2+1)$-dimensions along with the corresponding Lax
pair. For particular gauge choice this may result derivative and also
a higher order nonlinear
 extension of the Devay-Stewartson  equation (DSE),the
higher order DSE being a higher dimensional generalisation of the Kundu-
Eckhaus equation.
Such gauge transformation is
shown also to produce significant
extensions   to  the constrained
matrix KP system.
\end{abstract}
\\ \\ \\

Running title: {\it Derivative and higher order DS
 }
\\ \\ \\ \\ \\ \\ \\ \\ \\  \\

\hrule
{\it $^{\dagger}$Permanent address}:
 Saha Institute of Nuclear Physics, AF/1 Bidhan Nagar, Calcutta 700 064,
  India.

\newpage
%%%%%%%%%%%%%%%%%%%%%%%%%%%%%%%%%%%%%%%%%%%%%%%%%%%%%%%%%%%%%%%%%%
\section{Introduction}
\setcounter{equation}{0}
Though in $(1+1)$-dimensions there are a number
 of integrable systems solvable by inverse
 scattering transform  and allowing linear systems with Lax pair, in
 $(2+1)$-dimensions  such systems are surprisingly  very few in number.
  The well known integrable field equations
   in    $(2+1)$-dimensions  perhaps are
  only            the
   Kadomtsev-Petviashvili (KP) [1-3], the
    modified KP [4,5], the Ishimori
  [6]   and the Devay-Stewartson (DS) equation [7].
  While the KP and the modified KP may be looked as  higher dimensional
  generalisations of
   the KdV and the modified KdV equations,respectively  the
  DS equation
   is a generalisation of the nonlinear Schr\"odinger (NLS)
  equation. However  on the other hand,
   there exists a number of  integrable extensions of NLS
  equation in   $(1+1)$-dimensions, namely different varieties of
   derivative NLS [8,9] as well as  a higher order nonlinear extension
   of NLS known as
   Kundu-Eckhaus equation [10-12]. Therefore it is natural to ask whether
   such integrable extensions can also be generalised
   to higher  dimensions. Our aim here is
   precisely  to focus on this problem
   and show that  starting from a matrix KP [13-14]  it is possible
   to derive a   system of equations in $(2+1)$- dimensions along with
   its Lax pair representing extensions to the  DS equation.
Moreover  such a system would
contain as a subsystem the gauge transformed
  matrix KP hierarchy.
  Interestingly, for a particular gauge choise we
 obtain a new integrable equation in $(2+1)$-dimensions,
  which may be considered as a derivative DS  equation.
    On the other hand for a different choice of variables
   one gets yet another  higher order nonlinear extension of the DS
   equation, which  is indeed
    a $(2+1)$-dimansional                    integrable
     generalisation of the Kundu-Eckhaus equation and
   recovers the same when reduced to  $(1+1)$-dimensions.
    Such gauge transformations are found also to be compatible with the
   constrained matrix KP system and
   are able to   yield new integrable  systems in $(1+1)$-dimensions.

The paper is organised as follows. Sec 2. gives general framework of the
matrix KP hierarchy and shows how the DS equation fits into such a system.
Sec 3. derives some general  integrable extensions to the DS system within the
set up of matrix KP. In sec 4. we construct gauge transformed systems and
obtain the derivative DS as well as higher order nolinear DS equation
. Sec. 5
describes the gauge transformation of constrained matrix KP system
and finds the corresponding integrable equations. Sec. 6  is the concluding
section.

\section{ Preliminaries of matrix KP hierarchy and derivation of
DS equation
 }
\setcounter{equation}{0}
The matrix KP hierachy may be formulated [13-14] in analogy with the standard
(scalar) KP system  with  the difference that
 the basic pseudo-differential operator
\begin{equation}
W=1+ \omega_1 \partial^{-1}+ \omega_2 \partial^{-2} + \cdots
\end{equation}
%(2.1)
are now with  the matrix valued coefficients  $\omega_i$ .
That is   the dynamical
fields of the matrix KP hierarchy $ \omega = \{ \omega_i \}
\in g_0$ , where $ g_0$ is a finite dimensional algebra.
One may also invert the operator (2.1) to get
\begin {equation}
W ^{-1}=1- \omega_1 \partial^{-1}- ( \omega_2 -  \omega_1^{2}) \partial^{-2} +
\cdots
.\end {equation}
%(2.2)
For defining now the dynamics of the field $\omega_i$ along certain
`direction' $t_C$ we
 may choose a differential operator with matrix coefficients
as   $C= \sum_n a_n \partial^n, \ n \in {\bf N}, a_n \in g_0$ and
$[a_n , a_m] = 0$ and  require
\begin {equation}
W_{t_C}=-P_{< 0}(WCW^{-1}) W=-WC+M_CW
,\end {equation}
%(2.3)
where
\[
M_{C}= P_{\geq 0}(WCW^{-1})  \,.
\]
If $O$ is a pseudo-differential operator then we define its positive
(purely differential) part by $P_{0>}(O)$ and the
 negative part by $P_{<0}(O)$,
while the zeroth order terms of $O$ are denoted by $P_0(O)$.
 One can define as well a Lax operator
 $ L= W C W^{-1}$ and obtain
from (2.3) the Lax evolution equation in the standard form
\begin {equation}
L_{t_C}= [M_{C}, L ].
\end {equation}
%(2.4)
%
Differentiating (2.3) with respect to two different times $t_A$ and $t_B$
one arrives at the condition
\begin {equation}
  (W_{t_A})_{t_B} - (W_{t_B})_{t_A} = -P_{< 0}(W(A_{t_B}-B_{t_A}+[A,B])
  W^{-1}) W
,\end{equation}
%(2.5)
 or         \begin{equation}
      M_{A,t_B}-  M_{B,t_A} +
  [M_{A}, M_{B}] =  P_{\geq 0}(W(A_{t_B}-B_{t_A}+[A,B])
  W^{-1})
.\end {equation}
%(2.6)
It is evident that when the {\it rhs} of (2.5-6) vanishes one gets the usual
integrability condition
   \begin {equation}
  (W_{t_A})_{t_B} - (W_{t_B})_{t_A} =   M_{A,t_B}-  M_{B,t_A} +
  [M_{A}, M_{B}] =0,
\end {equation}
%(2.7)
which we shall use actively for formulating our dynamical system.
This may be achieved obviously by demanding simply the coefficients $a_n$
of the time-evolution operators $A$ and $B$ to be commuting as well as
 constant matrices giving $A_{t_B}=B_{t_A}=[A,B]=0$.  The usual practice
 [14] is to confine the choice $a_n \in(I,\sigma_3)$
 , which is also the case with the standard DS equation. However it is
important
 to note that (2.7) may be  fulfilled under more general  condition
            \begin{equation}
  A_{t_B}-B_{t_A}+[A,B]=0
,\end {equation}
%(2.8)
 which is itself like an integrability condition. As we will see below
 our proposed extension uses such $A,B$ operators
  with time-dependent coefficients, which satisfiy
 the more general integrability condition (2.8).
For the choice  $C=  c \partial^N $, we may get the Lax operator
\begin {eqnarray}
 L &=& W c \partial^N W^{-1} \nonumber \\
  &=&  (1+ \omega_1 \partial^{-1}+  \cdots )
  c \partial^N(
 1- \omega_1 \partial^{-1}-\cdots)  \nonumber \\ &=&
  c \partial^N +[\omega_1, c]
   \partial^{N-1} + u_{N-2}  \partial^{N-2} +\cdots
  +u_0+   u_{-1}  \partial^{-1}  +\cdots
,\end{eqnarray}
%(2.9)
where all coefficients $u_i$ can be expressed as differential
 expressions of $\omega_i$'s . From (2.3) we  also obtain
\begin {equation}
\omega_{1,t_C}=  u_{-1}   \ \ \mbox{and} \  \ M_C= c \partial^{N}
+[\omega_{1},c]  \partial^{N-1}+ \cdots +  u_{0} .
\end {equation}
%(2.10)
Now to see how the DS equation can be constructed from this set up
as a dynamical system, we restrict to  $(2\times2)$
matrices for $\omega_i$  and choose $A=\sigma_3 \partial$ and
       $B=\sigma_3 \partial^2$
for the evolution in the $y$ and $t$ directions, respectively [14].
{}From (2.1-3)  we derive
\begin {equation}
   W A W^{-1} =    \sigma_3 \partial
+[\omega_{1}, \sigma_3] + X \partial^{-1}+ ( \ ) \partial^{-2}+\cdots
,\end {equation}
%(2.11)
\begin {equation}
   W B W^{-1} =    \sigma_3 \partial^2
+[\omega_{1}, \sigma_3] \partial^{}+ ( X-\sigma_3 \omega_{1x}) +( \ )
 \partial^{-1}+\cdots
\end {equation}
%(2.12)
and
    \begin {equation}
 \omega_{1y} = - X
,\end {equation}
%(2.13)
where
\begin {equation}
X= [\omega_{2},\sigma_3  ]-\sigma_3 \omega_{1x} -[\omega_{1},\sigma_3]
 \omega_{1}
.\end {equation}
%(2.14)
Moreover
from (2.11-12) we clearly obtain
         \begin {equation}
M_{A}=   \sigma_3 \partial + U,
\end {equation}
%(2.15)
and
 \begin {equation}
M_{B}=   \sigma_3 \partial^{2}+U\partial + \frac{1}{2}(  \sigma_3 U_y+U_x)
+  \frac{1}{2}(D_y+  \sigma_3 D_x ),
\end {equation}
%(2.16)
where $\qquad \omega_1=-\frac{1}{2}(D+  \sigma_3 U)
 \qquad $ with
     $$
U= \left( \begin{array}{c} 0 \ \  q \\ r \ \  0
  \end{array}\right)=[\omega_{1},\sigma_3]
= -2\sigma_3 \omega_{1}^{off},
   \mbox{ \qquad   and \qquad}
    D= \left( \begin{array}{c} D_1  {\quad} \\ {\quad }  D_2
     \end{array}\right)
= -2 \omega_{1}^{diag}  $$
    matrices are introduced.
It is not
 difficult to check  that (2.15-16) represent the Zakharov-Shabat operators
for the DS system and the corresponding compatibility
 condition (2.7) leads to
the DS1 [7] equation
 \begin {equation}
  U_t=  \frac{1}{2}  \sigma_3 (U_{xx} + (U_{yy} )- \sigma_3  U^{3}
  + \sigma_3  \{D_x, U\}_+
\end {equation}
%(2.17)
for the {\it off-}diagonal elements with $\{ \ , \ \}_+ $
 denoting anti-commutators
, along with the relation
 \begin {equation}
    \sigma_3 D_y-  D_x = -U^2
\end {equation}
%(2.18)
obtained from the {\it diagonal} part of the relation (2.13).
Note also that due to condition (2.18) the  {\it diagonal}  elements
of the equation (2.7)     $$
D_{yy} -D_{xx}+(U^2)_{x}+\sigma_{3}(U^2)_{y}=0  $$
vanishes identically.

Introducing {\it light-cone} coordinates
     $\quad u=\frac{1}{2}(x+y),
 \ \ v=\frac{1}{2}(x-y)\quad
   $   and the notation
$\qquad A=\frac{1}{2}(D_y+  \sigma_3 D_x)
 \qquad $
one can rewrite DS equation  (2.17-18) in a more convenient form
as \begin {equation}
   U_t=    \sigma_3 (U_{uu} + (U_{vv} )
   +[A,U]
\end {equation}
%(2.19)
with  $A= diag (A_1,A_2)$ and
$$    A_{1u}=-  \frac{1}{2}(   U^2 )_{v} , \qquad  A_{2v}=
 \frac{1}{2}(   U^2 )_{u}
 . $$
\section{Extension of DS equation allowed by the matrix KP hierarchy }
\setcounter{equation}{0}

It is interesting to observe that in the framework of the matrix KP hierarchy
described above, it is posssible to extend the DS equation to a new
system of equations integrable in $(2+1)$ dimensions. This may be achieved
  by
 choosing the $y$ and $t$-directional
operators $A,$  and $B$   in a more general form
 \begin {equation}
A=   \sigma_3 \partial+  a_0
\end {equation}
%(3.1)
and  \begin {equation}
{B}=   \sigma_3 \partial^{2}+b_1 \partial +
b_0
,\end {equation}
%(3.2)
  where $ a_0, b_1,$ and $ b_0$ are
   now diagonal matrices with $(x,y,t)$
  dependent  elements and $A,B$ operators satisfy more relaxed condition
  (2.8).
    It is evident that for vanishing
   $ a_0, b_1,$ and $ b_0$
  the $A,B$ operators turn into those of the DS system. Now for
  finding the corresponding dynamical system,   using (2.3)
  and (2.10) we derive
    \begin {equation}
  M_{A}=   \sigma_3 \partial + U +a_0,
\end {equation}
%(3.3)
and \begin {equation}
M_{B}=   \sigma_3 \partial^{2}+(U+b_1) \partial - \omega_{1y}-\sigma_3
\omega_{1x}
+  b_0 - [\omega_1, (a_0-b_1)]
\end {equation}
%(3.4)
     along with
 \begin {equation}
      \omega_{1y}=-(X+ [\omega_1, a_0]),
\end {equation}
%(3.5)
with $X$ as in (2.14),
which  may be compared  again with the DS system (2.15-16).
As it has been mentioned already, for the validity
of the compatibility condition (2.7) the general necessary condition
(2.8) must be satisfied, which for our operators $A , B$ give the
condition
 \begin {equation}
 b_{1x} -\sigma_3  b_{1y} -2  a_{0x}=0
\end {equation}
%(3.6)
and \begin {equation}
 a_{0t} - b_{0y}+\sigma_3  b_{0x} - \sigma_3  a_{0xx}-b_1 a_{0x }=0
.\end {equation}
%(3.7)
The condition (3.5) on  the other hand gives  the same DS relation
$ \
      \sigma_3 D_y-  D_x = -U^2
\ .$
 It is interesting to note that using the DS relation together
with the necessary condition  (3.6-7) one obtains now from the
{\it diagonal} part of (2.7) an extra condition
  \begin {equation}
     [U , (a_0-b_1)]  =0
,\end {equation}
%(3.8)
       forcing  the relation
     $a_0-b_1= a{\bf I}$ with $a$ as a scalar function
      and as a consequence  the part
     $\quad   [\omega_1 , (a_0-b_1)]
     \quad$ dropps out from   expression (3.4) of $M_B$ .
 The {\it off-diagonal} part in its turn describes  the extended
 DS equation as
  \begin {equation}
    U_t=  \frac{1}{2}  \sigma_3 (U_{xx} + U_{yy} ) +\frac{1}{2}
[A,U ]+ [b_0,U] -\frac{1}{2} [a_0,\sigma_3 U_y+U_x]+
Ua_{0x} +   b_1  U_{x}.
\end {equation}
%(3.9)
  Therefore the set  (3.6-9) represents an integrable system of equations
  in $(2+1)$ dimansions given by Zakharov-Shabat operators
   $\ M_A, M_B \ $ as (3.3-4).
\section{Gauge
 transformation  leading to derivative and higher-order DS equations
}
\setcounter{equation}{0}
 Consider a gauge transformation by diagonal matrices $\theta$ over the basic
 operators $W$ : $$ \tilde W= e^{-\theta}W, \ \  \tilde W^{-1}=W^{-1}
  e^{\theta}
 $$
by  defining  new field variables  $\quad \tilde \omega_j=
   e^{-\theta}
  \omega_j e^{\theta}         \quad $
 consistent with such transformation. We start from  the same
   $A, B$ operators
 of the DS system and gauge transform them to get
\begin{equation}
       \tilde   W A \tilde W^{-1} =    \sigma_3 \partial
+( \tilde U+ \sigma_3 \theta_x) + ( \tilde X + \sigma_3 [\tilde \omega_1 ,
\theta_x]) \partial^{-1}+ ( \ ) \partial^{-2}+\cdots
\end {equation}
%(4.1)
and
\begin {eqnarray}
  \tilde   W B  \tilde  W^{-1} =    \sigma_3 \partial^2
&+& ([ \tilde \omega_{1}, \sigma_3]+ 2 \sigma_3 \theta_x) \partial^{}
- \tilde \omega_{1y} -\sigma_3 \omega_{1x}
 +\sigma_3 [\tilde \omega_{1}, \theta_x]+ [\tilde \omega_{1},
 \sigma_3]\theta_x \nonumber \\ &+&  \sigma_3 (\theta_{xx}+\theta_{x}^2)+
   [\tilde \omega_{1}, \theta_y]
 +( \ )
 \partial^{-1}+\cdots
\end {eqnarray}
%(4.2)
where  $$
 \tilde X \equiv X(\tilde \omega_{j}) \ \mbox{and} \
\tilde \omega_{1y}=-(\tilde X+ \sigma_3 [\tilde \omega_{1}, \theta_x])
+  [\tilde \omega_{1}, \theta_y]. $$
Therefore the corresponding gauge transformed $y$ and $t$ evolution
operators are
  \begin {equation}
\tilde M_A=     e^{-\theta } \tilde M_A e^{\theta}   +
             (  e^{-\theta })_y  e^{\theta} =
          \sigma_3 \partial
+( \tilde U+ \sigma_3 \theta_x)- \theta_y
\end {equation}
%(4.3)
and
  \begin {eqnarray}
  \tilde M_B &=&     e^{-\theta } \tilde M_B e^{\theta}   +
             (  e^{-\theta })_t e^{\theta} \nonumber \\ &=&
              \sigma_3 \partial^2
+ ( \tilde U+ 2 \sigma_3 \theta_x) \partial^{}
- \tilde \omega_{1y} -\sigma_3 \omega_{1x}  \nonumber \\  &\qquad +&
     (-\theta_t+  \sigma_3 (\theta_{xx}+\theta_{x}^2))
  + [\tilde \omega_{1}, \theta_y+ \sigma_3\theta_x ]
.\end {eqnarray}
%(4.4)
  For finding  the field equation
 from  compatibility condition (2.7) one may repeat again
  for  operators (4.3-4)   the steps described above .
 Interestingly however, we may avoid this tedious  task, if we compare
 the forms (4.3-4) with (3.3-4) and notice their equality  for the
choice
  \begin {equation}
  a_0=-\theta_y+\sigma_3 \theta_x ,
\end {equation}
%(4.5)
  \begin {equation}
          b_1= 2 \sigma_3 \theta_x
\end {equation}
%(4.6)
and  \begin {equation}
b_0= -\theta_t+  \sigma_3 (\theta_{xx}+\theta_{x}^2).
\end {equation}
%(4.7)
Moreover it is encouraging to find that this particular choice
(4.5-7)  satisfies  eqns. (3.6) and (3.7) trivially
and thus fulfils also the necessary condition (2.8),
 yielding finally the gauge equivalent DS equation as (3.9)  with
the substitution (4.5-7) together with  $\quad
  a_0 - b_1= \theta_y+\sigma_3 \theta_x = a {\bf I}
$ and
   \begin {equation}
    \sigma_3 D_y-  D_x = - \tilde U^2
.\end {equation}
%(4.8)
For obtaining now  physically interesting models we make the
choice of gauge as
 \begin {equation}
\theta_1=-D_2, \ \theta_2= D_1, \ \ \mbox {where} \ \ \theta= diag
 (\theta_1,\theta_2).
\end {equation}
%(4.9)
This immediately fixes from (4.8)
the relation
\begin {equation}
 \theta_y+\sigma_3 \theta_x = -\tilde
 U^2
 \end {equation}
%(4.10)
 identifying $a=(\tilde q \tilde r)$.
Therefore a gauge transformed DS system may be given as
  \begin {eqnarray}
 \tilde U_t=  \frac{1}{2}  \sigma_3 &(& \tilde U_{xx} + \tilde U_{yy} )-
  \sigma_3  \tilde U^{3}
  + \sigma_3  \{D_x, \tilde U\}_+ \nonumber \\& + &
    [( -\theta_t+  \sigma_3 (\theta_{xx}+\theta_{x}^2),\tilde U] -
      \{\theta_x, \tilde U_y\}_+ +\sigma_3 [\theta_x, \tilde U_x]
        \nonumber \\ &-&
    2 \sigma_3 \tilde U \theta_{xx}+ \tilde U (\tilde U^2)_x
\end {eqnarray}
%(4.11)
together with the relations (4.8) and (4.10).
Comparing  with (2.17) the difference of (4.11) from the standard DS
 equation  becomes evident.
To see the structure of the additional terms more explicitly we may rewrite
 equation (4.11) in light-cone coordinates $u,v$ taking into account
the useful identities for a diagonal matrix $ \theta = diag (  \theta_1
,  \theta_2)$ as
$$
% \begin {eqnarray}
[ \theta, \tilde U]=\sigma_3(\theta_1- \theta_2) \tilde U,
 \  \{\theta, \tilde U\}_+=(\theta_1+ \theta_2) \tilde U, \
[ \sigma_3\theta, \tilde U]=\sigma_3(\theta_1+ \theta_2) \tilde U,
%  \begin {equation}
  $$ and the relation $ \theta_1 =-D_2, \ \theta_2= D_1 $ with
  \begin{eqnarray}
  \theta_{1u}&=&-      \theta_{2v}= -\frac{1}{2}(\tilde q \tilde r)
 ,\\
         \theta_{2u}&=&  \frac{1}{2}\int^v (\tilde q \tilde r)_u dv', \ \
          \theta_{1v}= - \frac{1}{2}\int^u (\tilde q \tilde r)_v du',
. \end{eqnarray}
%(4.12-13)
For finding the explicit form of $\theta_t$ we may time-differentiate
 the
 relations  (4.12) and using equation (4.11) calculate $
  (\tilde q \tilde r)_t$, which yields
  \begin{eqnarray}
   \theta_{1t} &=& - \frac{1}{2}(\tilde j^{(u)}+\int^u (\tilde j^{(v)}_{,v}
   +\tilde j^0)du' ),
    \\  \theta_{2t} &=&
     \frac{1}{2}(\tilde j^{(v)}+\int^v (\tilde j^{(u)}_{,u} +\tilde j^0)dv')
   ,\end{eqnarray}
   %(4.14-15)
   where
   \begin {eqnarray}
   \tilde j^0&=& -2 (\tilde q \tilde r)^2(( \theta_1+ \theta_2)_{uu}
    -( \theta_1+ \theta_2)_{vv}), \\
 \tilde   j^{(u)}&=& \tilde i^{(u)}+  (\tilde q \tilde r)^2
 -2 (\tilde q \tilde r)  \theta_{2u}
\end {eqnarray}
%(4.16-7)
with $ \ \tilde i^{(u)}= (\tilde q_u \tilde r-  \tilde q \tilde r_u)
\ $  and a similar expression for
$\tilde j^{(v)} $. Substitution of these relations leads (4.11)
 finally to the explicit form
\begin{eqnarray}
 \tilde U_t=   \sigma_3 ( \tilde U_{uu} + \tilde U_{vv} )
  & + &  \frac{3}{2} \tilde U ((\tilde U^2)_u + (\tilde U^2)_v)-
       \tilde U^2 (\tilde U_u + \tilde U_v)+\sigma_3\tilde U^5 \nonumber
       \\ &+&   \frac{\sigma_3}{2} (\tilde i^{(u)}+ \tilde i^{(v)})\tilde U +
(Nonl)
 ,\end {eqnarray}
%(4.18)
with the nonlocal additional terms given by
   \begin {eqnarray}
(Nonl)& = &        {\sigma_3} (\theta_{2u}- \theta_{1v} )\tilde U
-2 (\theta_{2u} \tilde U_u -\theta_{1v} \tilde U_v)
   + {\sigma_3} \left((\theta_{2u})^2+(\theta_{1v} )^2\right)\tilde U
    \nonumber  \\ &-&  {\sigma_3}(\theta_{2u} + \theta_{1v}) \tilde U^3 +
    ( \theta_{2uu}+\theta_{1vv} )\tilde U
\nonumber \\ &+&
 \frac{\sigma_3}{2} (\int^u\tilde j_{,v}^{(v)} du' +
  \int^v\tilde j_{,u}^{(u)} dv'  +
   \int^u j^{0} du'  +   \int^v j^{0} dv' )
\tilde U
.\end {eqnarray}
%(4.19)
  It is easy to conclude now that
 equations (4.18-19) together with the expressions (4.12-17) represent
a new extended DS equation with fifth-order nonlinear terms.
 Noteworthyly  under reduction to $(1+1)$-dimensions , when $x$-
dependence may be ignored, the gauge choice becomes simply as
 $\quad \theta_1=-\theta_2\equiv h \ $
 and  (4.11)
    reduces to a gauge transformed equation
  in $(t,y)$ variables only as
  \begin {equation}
\tilde q_t=  \tilde q_{yy} +2 (\tilde q \tilde r) \tilde q +
   2 (h_{yy}+2 h_{y}^2 -h_{t} ) \tilde q+4h_{y} \tilde q_y
  \end {equation}
%(4.20)
along with a similar equation for $\tilde r$.
 Using further a consistent gauge choice
 $$ h_y=\frac{1}{2}(\tilde q \tilde r) ,\ \
 h_t=\frac{1}{2}(\tilde q_y \tilde r -\tilde q \tilde r_y)+
  (\tilde q \tilde r)^2
  $$ etc. [10]  one gets from (4.20) an equation in the form
     \begin {equation}
\tilde q_t=  \tilde q_{yy} +2 (\tilde q \tilde r) \tilde q  +
   2 (\tilde q \tilde r)_y \tilde q- (\tilde q \tilde r)^2 \tilde q
\end {equation}
%(4.21)
 and similarly for $\tilde r$. It is  readily seen  that  (4.21)
 is the same  higher-order NLS equation called  Kundu-Eckhaus equation
 [10-12].
Therefore the $(2+1)$-dimensional higher-order
 DS equation given by (4.11) or (4.18-19)
together with their complementary relations may be considered as the
higher-dimensional generalisation of the $ (1+1)$-dimensional
 Kundu-Eckhaus  equation.
 However as we describe below,
 the higher-dimensional generalisation
exhibits  some novel feature not present
 in  lower dimensions.

                   It is known that the Kundu-Eckhaus equation
                    can be transformed back
                   into the NLS equation by a suitable  change  of
                   variable ($S$-integrability in the language of [15]):
  $\ \tilde q \rightarrow q= e^{2h}\tilde q$ .
   Therefore one might expect to have similar situation
  also in  its higher-dimensional generalisation i.e. in
the   higher-order DS equation (4.11). However,  we find  remarkably
  that under a variable change
 $ \tilde U \rightarrow U= e^\theta \tilde U e^{-\theta}$
  (4.11) does not reduce back to the original DS equation (2.17)
 , but instead yields
  a completely new equation given by
    \begin {equation}
 U_t=  \frac{1}{2}  \sigma_3 ( U_{xx} +  U_{yy} )-
  \sigma_3   U^{3}
  + \sigma_3  \{D_x,  U\}_+  +  U ( U^2)_x
.\end {equation}
%(4.22)
  Note that though (4.22) is much simplified
compared to    (4.11), it is  different from the standard
 DS1 equation given as (2.17).
 The explicit  difference
 is due to the presence of an additional term containing derivative
of the field and therefore this new integrable equation (4.22) in
 $(2+1)$-dimensions
may be named as the
 derivative DS (DDS) equation.  In $(1+1)$-dimensions
however, when the $x$-dependence vanishes the extra derivative term
also disappears and (4.22) reduces to the same NLS equation
 as obtained from the standard
DS equation. Thus
  such derivative extensions are linked basically with the properties
  of  higher dimensions.
  \section { Constrained matrix KP and its gauge equivalent
   integrable system}
\setcounter{equation}{0}
It is known that [14,16-18] if $\quad\Phi_i, \ \Psi_i, \ i = 1,\cdots m
\quad$ be the eigenfunctions and
adjoint eigenfunctions of the matrix KP hierarchy satisfying:
\begin {equation}
 \Phi_{it_A}=P_0(M_A \Phi_i)
, \qquad  \Psi_{it_A}=-P_0(M_A^*\Psi_i)
,\end {equation}
%(5.1)
then the evolution of $W$ given by
    \begin {equation}
 W_{z}=-\sum_{i=1}^m (\Phi_i \partial^{-1}  \Psi_{i}^\dagger)W
\end {equation}
%(5.2)
is compatible with the KP hierarchy. Therefore for  DS case (2.11) one
may use the constrained system
    \begin {equation}
    WAW^{-1}=\sigma_3\partial+U+ \sum_{i=1}^m \Phi_i \partial^{-1}
     \Psi_{i}^\dagger
\end {equation}
%(5.3)
giving in place of (2.14) the expression $\quad X=
 \sum_{i=1}^m \Phi_i
     \Psi_{i}^\dagger$ , which from relation (2.13)  using (2.18)
     leads
   to  the constraints
     \begin {equation}
U_y=2\sigma_3  \sum_{i=1}^m (\Phi_i
     \Psi_{i}^\dagger)^{off}, \qquad D_y=2
               \sum_{i=1}^m (\Phi_i
     \Psi_{i}^\dagger)^{diag}
\end {equation}
%(5.4)
     and
         \begin {equation}
D_x=2 \sigma_3 \sum_{i=1}^m (\Phi_i \Psi_{i}^\dagger)^{diag} +U^2
.\end {equation}
%(5.5)
By using these constraints  it is possible therefore to get rid of
the $y$ derivatives of the fields $U$ and $D$ expressing them through
$\Phi_i$ and  $\Psi_i$ , and consequently obtain an integrable set of equations
as the constrained DS system involving
$\quad  U,\Phi_i$ and $   \Psi_{i}$.
 Our  intention   here is to find
 an  integrable sytem, which may be
obtained from the constrained DS under the gauge transformation considered
in the last section.

We observe that under the gauge transformation with diagonal matrix $\theta$
we get the corresponding transformed objects as
    \begin {equation}
\tilde \Phi_i=e^{-\theta} \Phi_{i} ,\ \Psi_i^\dagger= \Psi_{i}^\dagger
 e^{\theta} , \ \ \mbox {and} \ \  \tilde U=e^{-\theta} U e^{\theta}
,\end {equation}
%(5.6)
while for defining their $y$ and $t$-evolutions
 we should use in (5.1) the expressions of
  $\tilde M_A$ and $\tilde M_B$ as given in
(4.3-4). Subsequently  the constraint may be imposed,which
would result
 emergence of $(1+1)$-dimensional systems for each pair of variables.
         Choosing now  the gauge functions
$\quad \theta= diag ( \theta_1, \theta_2)$  as before :$
 \  \theta_1=-D_2,   \
  \theta_2= D_1 \quad$
  one  gets  the explicit
    relation
 \begin{equation}
 diag(\theta_{2y}, - \theta_{1y}) =2
               \sum_{i=1}^m (\tilde\Phi_i
  \tilde   \Psi_{i}^\dagger)^{diag}
 \ \ \mbox{with} \  \
  \theta_y+\sigma_3\theta_x=-\tilde U^2,
\end {equation}
%(5.7)
 while the gauge transformed constrained may be given similar to (5.4) by
     \begin {equation}
\tilde U_y=2\sigma_3  \sum_{i=1}^m (\tilde \Phi_i
  \tilde   \Psi_{i}^\dagger)^{off}, \qquad D_y=2
               \sum_{i=1}^m (\tilde\Phi_i
  \tilde   \Psi_{i}^\dagger)^{diag}, \ \
    \sigma_3 D_y-  D_x = - \tilde U^2
.\end {equation}
%(5.8)
These constraints as mentioned before, eliminate the involvement
 of $y$-derivatives
from the system
 resulting finally   the set of gauge transformed constrained DS
 system  expresssed  in $\quad\tilde
  U, \tilde \Phi_i ,\tilde \Psi_{i}\quad$ . Thus    \begin {equation}
\tilde \Phi_{iy}=\sigma_3\tilde \Phi_{ix} +(\tilde U+\tilde U^2+ 2\sigma_3
                          {\theta_x})\tilde \Phi_i
,\end {equation}
%(5.9)
    \begin {equation}
\tilde \Psi_{iy}=\sigma_3\tilde \Psi_{ix} -(\tilde U^\dagger+
 \tilde U^{\dagger 2}+2\sigma_3
                          {\theta^\dagger_x})\tilde \Psi_i
\end {equation}
%(5.10)
together  with the realtions (5.7) and
constraint (5-8) describe the $y$-evolution, while
 for the $t$-evolution we
get
    \begin {eqnarray}
\tilde \Phi_{it}=\sigma_3\tilde \Phi_{ixx} &+&\frac{1}{2}
(\tilde U_x +\sigma_3 \tilde U^2)\Phi_i+(\tilde U
+ 2\sigma_3 {\theta_x})\tilde \Phi_{ix}+b_0 \Phi_i\nonumber \\& + &
   \sum_{j=1}^m (\tilde\Phi_j
  \tilde   \Psi_{j}^\dagger)\tilde \Phi_i+   \sum_{j=1}^m (\tilde\Phi_j
  \tilde   \Psi_{j}^\dagger)^{diag} \tilde \Phi_i ,
,\end {eqnarray}
%(5.11)
\begin{eqnarray}
  \tilde U_t=  \frac{1}{2}  \sigma_3 & & \tilde U_{xx}
 + \sigma_3  \tilde U^{3}
+    [b_0,\tilde U] -
     2 \{\theta_x,
\sigma_3  \sum_{i=1}^m (\tilde \Phi_i
  \tilde   \Psi_{i}^\dagger)^{off}
  \}_+
+ \sigma_3 [\theta_x, \tilde U_x]
        \nonumber \\ &-&
   2 \sigma_3 \tilde U \theta_{xx}+ \tilde U (\tilde U^2)_x+[2\sigma_3
  \theta_x, \sum_{i=1}^m (\tilde \Phi_i
  \tilde   \Psi_{i}^\dagger)^{off} ]         \nonumber \\& + &
\sigma_3 \sum_{i=1}^m (\tilde \Phi_{ix}
  \tilde   \Psi_{i}^\dagger -
         \tilde \Phi_i
  \tilde   \Psi_{ix}^\dagger)^{off}+
                          [    \sum_{i=1}^m (\tilde \Phi_i
  \tilde   \Psi_{i}^\dagger)^{diag},\tilde U]
,\end {eqnarray}
%(5.12)
and   similarly by an
equation for $\tilde \Psi_i$,
where $b_0= -\theta_t+\sigma_3( \theta_{xx}+\theta_x^2)$
. Note that
  $\theta_t$ appearing in the above equations can
be calculated in explicit form using (5.7) and taking into
 accout the time-evolution equations
 like (5.11-12). It is evident therefore that
  the gauge transformed
 constraint DS system
 (5.8-10) and (5.11-12) are
$(1+1)$-dimensional systems
 defined in variables
  $(y,x)$ and $(t,x)$, respectively and they represent
   new integrable
system of equations
         with higher-order nonlinearities.
\section{Concluding remarks }
Within the framework of matrix KP it is possible to construct integrable
 extensions to DS equation, which contains also gauge transformed
 systems.
Such extensions utilise effectively the more general freedom of choosing
coefficients of the time-evolution operators as functions of the
 time variables
instead of their standard choice as constants.
 Fixing suitably the gauge  we find a $(2+1)$-dimensional
 generalisation of Kundu-Eckhaus equation as well as a derivative DS
 equation. As far as we know these are new integrable systems in $(2+1)$-
 dimensions.
 Under reduction
  to  $(1+1)$- dimensions  the derivative DS goes back  to the
  same NLS equation as the standard DS equation, showing the derivative
   extension to be  essentially  an artifact of the higher dimensional
   generalisation. Such extended systems also preserve their features
   in the constrained case.
 \section*{ Acknowledgement}
One of the authors (AK) likes to express
 his thanks to the Alexander von Humboldt Foundation
for  the award of its research fellowship.

\newpage

\section*{References}
\begin{enumerate}
%1
\item Y.Ohta, J.Satsuma, D.Takahashi, T.Tokihiro, Prog. Theor. Phys.,
      Suppl., 94, 219 (1988).
%2
\item E.Date, M.Jimbo, M.Kashiwara, T.Miwa,
      in Nonlinear integrable systems --
      classical and quantum theory, ed. M.Jimbo, T.Miwa,
      World scientific, 39 (1983).\\
      M.Jimbo, T.Miwa, Publ.RIMS, Kyoto Univ., 19, 943 (1983).
%3
\item M.Sato, Y.Sato, in Nonlinear partial differential
      equations in applied sciences, ed. H. Fujita, P.D. Lax,
      G.Strang, Kinokuniya/North Holland, Tokyo/Amsterdam,
      259 (1983).
\item Yi Cheng , Yi-shen Li , J.Phys. A, 25, 419 (1992).
\item W.Oevel , W.Schief,
      Squared eigenfunctions of the (modified) KP hierarchy
      and scattering problems of Loewner type.
      Applied Mathematics Preprint AM 93/21,
      University of New South Wales, Australia (1993).

\item Y.Ishimori, Prog.Theor.Phys., 72, 33 (1984)
\item M.Boiti, J. Leon , F.Pompinelli, Phys.Lett.A, 141, 96 (1989).
 \item   H. H. Chen ,  Y. C. Lee and C.S. Liu, Phys.Scripta, 20 ,490 (1979).
\item D.J.Kaup , A.C.Newell, J.Math.Phys., 19, 789, (1978)
\item A.Kundu, J.Math.Phys., 25,  3433 (1984).
\item F.Calogero, Inverse Prob., 3, 229 (1987)
 \item Li Shen in 'Symmetries and
  singularity sructures' \quad (Springer Publication
, 1990,  ed. M. Lakshmanan), p.27

\item K.Kajiwara, J.Matsukidaira, J.Satsuma,
      Phys. Lett. A, 3, 115 (1990).
\item W.Oevel, Physica A, 195, 533 (1993).

\item F.Calogero , in 'What is Integrability' (Springer Publ.,1991,
ed. V.E.Zakharov ),p.1.
\item Y.Cheng, Y.S.Li, Phys. Lett. A, 157, 22 (1991).
\item B.G.Konopelchenko, J.Sidorenko, W.Strampp,
      Phys. Lett. A, 157, 17 (1991).
\item W.Oevel, W.Strampp,
      Commun. Math. Phys., 157, 51 (1993).

\end{enumerate}
\end{document}